\documentstyle[12pt,twoside]{article}

\begin{document}
{
\title{\Large\bf QUANTUM MODES AROUND A\\
SCALAR-TENSOR BLACK HOLE: BREAKDOWN\\
OF THE NORMALIZATION CONDITIONS}

\author{\Large F.G. Alvarenga, A.B. Batista,
J. C. Fabris and G.T. Marques\\
Departamento de F\'{\i}sica, Universidade Federal do Esp\'{\i}rito Santo, \\
CEP29060-900, Vit\'oria, Esp\'{\i}rito Santo, Brazil\\
} 
\maketitle

\begin{abstract}
Black holes arising in the context of scalar-tensor gravity theories, where
the scalar field is non-minimally coupled to the curvature term, have zero
surface gravity. Hence, it is generally stated that their Hawking temperature is
zero, irrespectivelly of their gravitational and scalar charges.
The proper analysis of the Hawking temperature requires to study the propagation
of quantum fields in the space-time determined by these objects. We study scalar fields
in the vicinity of the horizon of these black holes. It is shown that the scalar
modes do not form an orthonormal set. Hence, the Hilbert space is ill-definite in this
case, and no notion of temperature can be extracted for such objects.
\end{abstract}

The Hawking temperature is essentially a semi-classical effect \cite{hawking}. To obtain the notion
of temperature for a black hole, for example, the propagation of quantum fields must
be analysed in the dynamical process of formation of the black hole. The non-unicity of
the vacuum state during this process implies that the black hole
generates a continous flux of particles with a planckian spectrum \cite{birrel,spindel}. In order that this
analysis to be self-consistent, for each slice at constant time a complete set of orthonormal
modes must be defined, spanning a Hilbert space on such slices. Hence, even if the
computation of the Hawking temperature for a black hole requires the full analysis of
quantum fields during the black hole formation, considerable informations may be obtained 
by studying the quantum modes in a static configuration.
\par
In a previous work \cite{glauber1}, it has been argued that the notion of temperature for
any zero surface gravity black hole is ill-definite. The explicit analysis of the propagation
of quantum scalar fields for the Reissner-Nordstr\"om extreme black hole \cite{glauber2} indicates that
this is due to the failure of the semi-classical analysis: it is not possible to construct
a Hilbert space for scalar fields around an extreme black hole, at least when the horizon
is formed. This is due to the fact that the scalar modes, in spite of forming an orthogonal
set, have a divergent norm, which can not be normalized using the standard procedures: the distance of any point external to the event horizon to the event horizon itself is infinite,
hence any volume around the black hole is also infinite.
\par
In the present work we extend this analysis to black holes which arise
in the context of non-minimally coupled scalar-tensor theories \cite{kirill1,kirill3}. One important feature of
such black holes is that they all have zero surface gravity,
irrespectivelly of the value of the scalar charge, which tempt us to state
that they have zero temperature. However, it will be shown here that there is
no consistent Hilbert space, at least for scalar fields, in
the space-time generated by these holes. To do so, we will solve the Klein-Gordon equation for masseless fields
in the vicinity of the horizon. The quantum modes in this case,
as in the RN extreme one, form an orthogonal set,
but they do not constitute normalized states, neither they can be normalized using
the standard procedures. This is a quite surprising feature of the scalar-tensor black holes:
quantum field theory seems to be ill-definite in the space-time generated by these
objects. This rises the question if such object can really exist in nature.
\par
Let us first summarize the main properties of the scalar-tensor black holes.
For simplicity, let us consider those that come out from the Brans-Dicke theory,
defined by the Lagrangian density
\begin{equation}
\label{lag}
L = \sqrt{-g}\biggr(\phi R - \omega\frac{\phi_{;\rho}\phi^{;\rho}}{\phi}\biggl) \quad ,
\end{equation}
where $\omega$ is the usual (constant) Brans-Dicke parameter such that $\omega \neq -3/2$.
For $\omega = - 3/2$ the Lagrangian density (\ref{lag}) is conformally equivalent to
usual Einstein-Hilbert Lagrangian density. Solutions for the static spherically symmetric
space-time lead to the metric \cite{kirill1}
\begin{equation}
\label{m1}
ds^2 = \rho^{n+2}dt^2 - \frac{4k^2(m - n)^2}{(1 - P)^4}\rho^{-(n+2)}d\rho^2 - \frac{4k^2}{(1 - P)^2}\rho^{-m}d\Omega^2 \quad ,
\end{equation}
where
\begin{equation}
P \equiv 1 - \frac{2k}{r} \equiv \rho^{m - n} \quad .
\end{equation}
The event horizon appears at $\rho = 0$.
The parameters $m$ and $n$ are connected with the mass and the scalar charges. They must
be positive integers and must satisfy the relation $m - n \geq 2$, with $m, n > 0$,
in order that
the space-time defined by (\ref{m1}) admits a regular extension. The solutions imply
$\omega < - 3/2$.
When $m = 0$ and $n = -1$, the metric (\ref{m1}) describes the Schwarzschild solution.
Note that these black holes have an infinite horizon surface.
\par
Near the horizon, the metric (\ref{m1}) simplifies to
\begin{equation}
\label{m2}
ds^2 \approx \rho^{n+2}dt^2 - 4k^2(m - n)^2\rho^{-n-2}d\rho^2 - 4k^2\rho^{-m}d\Omega^2 \quad .
\end{equation}
The Klein-Gordon equation for a massless scalar field
\begin{equation}
\Box\Psi = 0 \quad ,
\end{equation}
in the vicinity of the horizon takes the form
\begin{equation}
\label{kge}
\psi'' - (m - n - 2)\frac{\psi'}{\rho} - \biggr\{(m - n)^2l(l + 1)\rho^{m - n - 2} - 4k^2(m - n)^2\omega\rho^{-2(n + 2)}\biggl\}\psi = 0 \quad ,
\end{equation}
where $\Psi(\rho,t) = \psi(\rho)e^{i\omega t}$.
\par
For the Reissner-Nordstr\"om case, this equation can be explicitly solved, leading to the usual
set of orthonormal modes \cite{matsas},
\begin{equation}
\Psi = c_1K_{i4\omega}\biggr(\frac{\sqrt{l(l+1)}}{r_+}\sqrt{\rho}\biggl)e^{i\omega t} \quad ,
\end{equation}
where $K_\nu(x)$ stands for the modified Bessel's function.
For the values of $n,m$ leading to the scalar-tensor black holes, equation (\ref{kge}) seems
to admit no closed solution for arbitrary value of $l$. However, for the monopolar case
$l = 0$, we can solve (\ref{kge}) explicitly. The solutions are
\begin{equation}
\label{s1}
\Psi(\rho,t) = \rho^{-(n + 1 - m)/2}\biggr\{c_1J_\nu(\tilde\omega\rho^{-(n+1)})
+ c_2J_{-\nu}(\tilde\omega\rho^{-(n+1)})\biggl\}e^{i\omega t}
\end{equation}
where $\nu = (n + 1 - m)/[2(n + 1)]$ and $\tilde\omega = 2k(m - n)\omega/(n + 1)$. From the
above solutions, the standard method of quantization can be applied, decomposing
the modes into creation and annihilation operators.
\par
The question we would like to address now is if the solutions (\ref{s1}) constitute a set of
orthonormal modes. The inner product associated with the Klein-Gordon equation
is defined as
\begin{equation}
(\Psi_1,\Psi_2) = - i\int d\Sigma^\mu\biggr(\Psi_1\partial_\mu\Psi^*_2 - \partial_\mu\Psi_1\Psi_2^*\biggl) = -i\int d\Sigma^\mu\Psi_1\stackrel{\leftrightarrow}{\partial}_\mu\Psi_2^* \quad ,
\end{equation}
where $\Sigma^\mu$ is a spacelike hypersurface. When this hypersurface is defined with
$t = $ constant, the expression for the inner product takes the form
\begin{equation}
(\Psi_1,\Psi_2) = -i\int dx^3\sqrt{-g}g^{00}\Psi_1\stackrel{\leftrightarrow}{\partial}_t\Psi_2^* \quad ,
\end{equation}
Applying this expression for the solutions found above, we obtain that the norm of the modes
are given by
\begin{equation}
(\Psi,\Psi) \propto \int_0^\infty x|J_{\pm\nu}(\tilde\omega x)|^2
dx \quad ,
\end{equation}
where $x = \rho^{-(n + 1)}$. Note that as the horizon is approached, $x \rightarrow \infty$.
The norm of these modes are infinite, as it can be easily seen by inspecting the asymptotic
behaviour near the horizon:
\begin{equation}
(\Psi,\Psi) \propto \int^\infty |\cos(\tilde\omega x - \delta)|^2
dx \quad ,
\end{equation}
where $\delta$ is a constant phase.
\par
The fact that the solutions found for the Klein-Gordon equation near the horizon do not
constitute a complete set of orthonormal modes is, in principle, not a problem. In flat
space-time, the solutions of the Klein-Gordon equation are given in terms of plane waves,
which are not also normalized states. But, usually this problem can be circumvented by
defining the plane waves modes in a finite volume. Hence, the notion of a Hilbert space
is recovered. At the end, when the calculations for the physical quantities
are performed, the infinite volume limit is taken leading to finite answer.
In fact, the inner product between two diferent modes found above obeys the well-known
Bessel's orthogonalization condition:
\begin{equation}
(\Psi_\omega,\Psi_{\omega'}) \propto \int_0^\infty xJ_{\pm\nu}(\tilde\omega x)J_{\pm\nu}(\tilde\omega'x)
dx = \delta_{\tilde\omega,\tilde\omega'}\quad .
\end{equation}
Hence, in principle we face the same situation as it happens with plane waves in
a flat space-time.
But, here the
problem is more complex. The reason is that the distance of any point to the event horizon
in the space-time defined by (\ref{m1}), with the conditions leading to the scalar-tensor black holes,
is infinite. In fact, taking the metric (\ref{m2}) and integrating the radial distance,
one obtain
\begin{equation}
l = 2k(m - n)\int_0^\rho\frac{d\rho'}{{\rho'}^{(n+2)/2}} \rightarrow \infty
\end{equation}
since $n \geq 0$. Hence, any volume around the horizon is infinite. The conclusion is that the
solutions of the Klein-Gordon equation in the scalar-tensor black hole case can not be normalized.
The same situation occurs in the RN extreme black hole case \cite{glauber1}.
\par
This result has, in our opinion, some dramatic consequences. The calculation presented
here should be extended for spinorial and vectorial modes, but we can expect similar conclusions.
The main point is that there is no Hilbert space in the space-time defined by the scalar-tensor black holes.
Hence, there is no way to implement consistently a quantum field theory in this space-time.
Normally, we could expect that in any reasonable space-time a quantum field theory can be
implemented, if we agree with the universality of general relativity and quantum mechanics.
Perhaps, this compatibility can be recovered, but it is not clear how.
In fact, some problems of consistency appear when quantum fields are considered in,
for example, an anti-deSitter space-time. This is due to the fact that the anti-deSitter
space-time is not globally hyperbolic \cite{ellis}. However, it has been shown \cite{isham} that
quantum field theory can be consistently formulated in an anti-deSitter space-time
if convenient boundary conditions are chosen in the timelike infinity. An application
of such procedure occurs, for example, with the AdS black holes \cite{witten,horowitz}, which play a central
r\^ole in the AdS/CFT correspondance \cite{petersen,douglas}.
\par
However, in the case of the scalar-tensor black holes, the situation is more involved. First
of all, it must be noticed that the near-horizon geometry is very similar to
the anti-deSitter geometry. But, there is no timelike infinity as in the
universal covering of the anti-deSitter space-time. Moreover, it seems that no
particular boundary condition can cure the problem treated here, since all solutions
of the Klein-Gordon equation have infinite norm. The problem seems to be very delicate
because it reveals that perhaps not all regular space-time is compatible with
quantum mechanics. Perhaps, there is some possible mechanism to circumvent this
anomaly, as it has been done in reference \cite{isham} for the anti-deSitter space-time.
But, at this moment it is far from being obvious which mechamism could give sense
to a quantum theory in the space-time of zero surface gravity black holes.
\par
The conclusion that no quantum field theory can be consistently implemented in the
space-time generated by the scalar-tensor black holes (or, more generally, by any
zero surface gravity black hole) may seem quite radical. However, this is in
agreement with the fact that the semi-classical analysis for these objects
is ill-definite: the Bogolubov's coefficients do not obey the required compatibility
conditions \cite{glauber1,glauber2}. Moreover, for the RN extreme black hole
it has already been shown that no notion of temperature can be obtained through the
euclideanization of metric, in opposition to what happens in
the non-extreme cases \cite{hawking1}: the periodicity of the euclidean time is arbitrary indicating
an arbitrary temperature. The entropy law $S = A/4$, $A$ being the
area of the event horizon, fails also for zero surface gravity black holes
\cite{hawking1,zaslavskii}. Combined with the analysis presented in
this note, all these results
indicates that quantum field theory in the space-time of zero surface gravity black holes
is not consistent. The comom features of all zero surface gravity black holes are:
the distance of any point to the event horizon is infinite; the event horizon
is degenerated, being also a Cauchy horizon. The failure to implement a quantum field theory
in the space-time of these objects may be, perhaps, searched with the aid of these
particular features of the zero surface gravity black holes.
\newline 
\vspace{0.5cm}
\newline
{\bf Acknowledgements:} This work has been
partially supported by CNPq (Brazil) and CAPES (Brazil).


\begin{thebibliography}{100}
\bibitem{hawking} S.W. Hawking, Comm. Math. Phys. {\bf 43}, 199(1975);
\bibitem{birrel} N.D. Birrell and P.C.W. Davies, {\bf Quantum
fields in curved space}, Cambridge University Press,
Cambridge(1982);
\bibitem{spindel} R. Brout, S. Massar, R. Parentani and Ph.
Spindel, Phys. Rep. {\bf 260}, 6(1995);
\bibitem{glauber1} F.G. Alvarenga, A.B. Batista, J.C. Fabris and
G.T. Marques, {\it Zero surface gravity black holes and the failure of semi-classical analysis},
gr-qc/0309022;
\bibitem{glauber2} F.G. Alvarenga, A.B. Batista, J.C. Fabris and G.T. Marques, Phys. Lett. {\bf A320}, 83(2003);
\bibitem{kirill1} K.A. Bronnikov, G. Cl\'ement, C.P. Constantinidis and J.C. Fabris,
Grav.\&Cosm. {\bf 4}, 128(1998));
\bibitem{kirill3} K.A. Bronnikov, G. Cl\'ement, C.P Constantinidis and J.C. Fabris,
Phys. Lett. {\bf A243}, 121(1998);
\bibitem{matsas} J. Castineiras, L.C.B. Crispino, G.E.A. Matsas and D.A.T. Vanzella,
Phys.Rev. {\bf D65}, 104019(2002);
\bibitem{ellis} S.W. Hawking and G.F.R. Ellis, {\bf The large scale
structure of space-time}, Cambridge University Press, Cambridge
(1973);
\bibitem{isham} S.J. Avis, C.J. Isham and D. Storey, Phys. Rev. {\bf D18}, 3565(1978);
\bibitem{witten} E. Witten, Adv. Theor. Math. Phys. {\bf 2}, 505(1999);
\bibitem{horowitz} G.T. Horowitz and S.F. Ross, JHEP {\bf 9804}, 15(1998);
\bibitem{petersen} J.L. Petersen, Int. J. Mod. Phys. {\bf A14}, 3597(1999);
\bibitem{douglas} M.R. Douglas and S. Randjbar-Daemi, {\it Two lectures on
AdS/CFT corerspondence}, hep-th/990202;
\bibitem{hawking1} S.W. Hawking, G.T. Horowitz S.F. Ross, Phys. Rev. {\bf D51}, 4302(1995);
\bibitem{zaslavskii}  O. B. Zaslavskii, Class. Quant. Grav. {\bf 19}, 3783(2002).

\end{thebibliography}
\end{document}